# DEFINABILITY IN PHYSICS

D.J. BENDANIEL
Cornell University
Ithaca NY, 14853, USA

**Abstract.** The concept of definability of quantum fields in a set-theoretical foundation is introduced. We propose an axiomatic set theory and then derive a nonlinear sigma model and the Schrödinger equation in a Lagrangian form; this follows naturally from a null postulate which expresses symmetry of action. Definability in this theory is necessary and sufficient for quantum mechanics. Space-time proves to be relational and the fields are free of singularities.

We look to provide a deep connection between physics and mathematics by requiring that quantum fields be definable in a set-theoretical foundation. For mathematics, the usual foundation is the set theory of Zermelo-Fraenkel (ZF). In ZF, a set U of finite integers is definable if and only if there exists a formula $\Phi_U(n)$ from which we can unequivocally determine whether a given finite integer n is a member of U or not. That is, when a set of finite integers is not definable, then there will be at least one finite integer for which it is impossible to determine whether it is in the set or not. Other sets are definable in a theory if and only if they can be mirrored by a definable set of finite integers. Most sets of finite integers in ZF are not definable. Furthermore, the set of definable sets of finite integers is itself not definable in ZF. [1]

A quantum field in a finite region of space is definable in a set-theoretical foundation if and only if the set of distributions of the field's energy among its eigenstates can be mirrored in the theory by a definable set of finite integers. This concept of definability is appropriate because, were there a field whose set of energy distribution among eigenstates corresponded to an undefinable set of finite integers, that field would have at least one energy distribution whose presence or absence is impossible to determine, so the field could not be verifiable or falsifiable, i.e., the existence of the field could not be empirically established. Therefore, our task is to find a foundation in which it is possible to specify completely the definable sets of finite integers and which also contains mathematics rich enough to obtain the quantum fields corresponding to these sets.

The definable sets of finite integers cannot be specified completely in ZF because there are infinitely many infinite sets whose definability is undecidable. So we must start with a sub-theory containing no infinite sets of finite integers. Then all sets of finite integers are *ipso facto* definable. This will mean, of course, that the set of all finite integers, called $\omega$, cannot exist in that sub-theory. The set $\omega$ exists in ZF essentially in consequence of two axioms, the axiom of infinity and the axiom schema of subsets. Thus, we must delete one or the other of these axioms. If we delete the axiom of infinity we will then have no need for the axiom schema of subsets either since all sets are finite. However that theory is not rich enough to obtain continuous functions of a real variable, so it is not useful for quantum mechanics. So the task reduces to whether or not, starting by deleting the axiom schema of subsets from ZF but retaining the axiom of infinity, we get a theory rich enough to obtain the quantum fields corresponding to its sets of finite integers.

In the appendix we show seven axioms. The first six are the axioms of ZF except that the axiom schema of replacement has been modified and the power set axiom has been deleted. The usual replacement axiom (AR) asserts that, for any functional relation, if the domain is a set, then the range is a set. That axiom actually combines two independent axioms: the axiom schema of subsets, which we wish to delete, and an axiom schema of bijective replacement (ABR), which refers only to a one-to-one functional relation. We can thus first delete the axiom schema of subsets from ZF by substituting ABR for AR, producing ZF – Subsets.

We shall then see how ZF – Subsets differs importantly from ZF by looking at the axiom of infinity. The axiom of infinity asserts the existence of at least one set $\omega^*$. There are actually infinitely many such sets. In ZF, we can obtain the minimal $\omega^*$, a set usually called $\omega$, by using the axiom schema of subsets to provide the intersection of all the sets created by the axiom of infinity. However, without the axiom schema of subsets the set $\omega$ cannot be obtained; all theorems of ZF – Subsets must hold for any $\omega^*$. Any member of $\omega^*$ is an "integer". An infinite integer is a member of $\omega^*$ that is one-to-one with $\omega^*$. A "finite integer" is a member of $\omega^*$ that is not infinite. In the subsequent discussion we denote finite integers by *i, j, k, ℓ, m* or *n*.

We now adjoin to ZF − Subsets another axiom asserting that all subsets of ω* are constructible. By constructible sets we mean sets that are generated sequentially by some process, one after the other, so that the process well-orders the sets. Gödel has shown that an axiom asserting that all sets are constructible can be consistently added to ZF, giving a theory called ZFC[+].[2] It has also been shown that no more than countably many subsets of ω* can be proven to exist in ZFC[+].[3] These results will, of course, hold for the sub-theory ZFC[+]− Subsets. Therefore we can adjoin to ZF − subsets a new axiom asserting that the subsets of ω* are countably constructible. If we in addition remove the power set axiom, then we are assured that non-constructible subsets of ω* do not exist. We call ZF − Subsets − Powers + Constructibility the theory T.

Cantor's proof or its equivalent is not available in T [4]; no uncountably infinite sets exist in T. Since all sets are countable, the general continuum hypothesis holds. Moreover, in the theory T, all sets of finite integers are finite, so any infinite set of integers must contain infinitely many infinite integers. Furthermore, we cannot show the induction theorem, so not even all the countable sets that exist in ZF are allowed in T. For example, we cannot sum infinite series, whereas in ZF infinite series play an important role in the development of mathematics. Nevertheless, our axiom of constructibility provides a novel route for obtaining functions of a real variable.

We start by showing that the theory T contains a real line. Recall the definition of "rational numbers" as the set of ratios, usually called **Q,** of any two members of the set ω. In T, we can likewise, using the axiom of unions, establish for ω* the set of ratios of any two of its integers, finite or infinite. This will be an "enlargement" of the rational numbers and we shall call this enlargement **Q\***. Two members of **Q\*** are called "identical" if their ratio is 1. We employ the symbol "≡" for "is identical to". An "infinitesimal" is a member of **Q\*** "equal" to 0, i.e., letting $y$ signify that member and now employing the symbol "=" to signify equality, $y = 0 \leftrightarrow \forall k[y < 1/k]$. The reciprocal of an infinitesimal is "infinite". Any member of **Q\*** that is not an infinitesimal and not infinite is "finite", $[y \neq 0 \wedge 1/y \neq 0] \leftrightarrow \exists k[1/k < y < k]$. We apply this concept of equality to the interval between two finite members of **Q\***; two finite members are either equal or the interval between them is finite. The constructibility axiom in T will now well-order the power set of ω*, creating a metric space composed of the subsets of ω*. These subsets represent binimals making up a real line **R***. [5] In this theory **R*** is a subset of **Q***.

Now *equality-preserving* bijective mappings between finite intervals of **R*** are homeomorphic, i.e., bijective mappings $\phi(x,u)$ of a finite interval $X$ onto a finite interval $U$ in which $x \in X$ and $u \in U$ such that $\forall x_1, x_2, u_1, u_2 [\phi(x_1, u_1) \wedge \phi(x_2, u_2) \rightarrow (x_1 - x_2 = 0 \leftrightarrow u_1 - u_2 = 0)]$ will produce biunique function pieces which are continuous when taken as either $u(x)$ or $x(u)$. Note that if $X$ or $U$ vanishes then the other must vanish as well. The derivatives of these pieces can be obtained using corresponding infinitesimals $du$ and $dx$. The pieces can, of course, be connected to obtain more functions of a real variable. $u(x)$ is a "function of a real variable in T" only if it is a constant (obtained directly from ABR) or a continuously connected sequence of biunique pieces such that its derivative $\left(\dfrac{du}{dx}\right)$ is also a function of a real variable in T. If some derivative is a constant, these conditions are necessary and sufficient to define polynomial functions. If no derivative is a constant, then strictly speaking those functions do not exist in T, since they require an infinite series description [6]. Nevertheless, we will incorporate such functions by representing them arbitrarily closely, quite analogous to Sturm-Liouville expansions of the usual mathematics, by a linear combination of polynomials of sufficiently high degree each of which is obtained iteratively from the following well-known integral expression and boundary conditions:

$$\int_a^b [p\left(\dfrac{du}{dx}\right)^2 - qu^2] dx \equiv \lambda \int_a^b ru^2 dx ; \ \lambda \text{ is locally minimum for } \int_a^b ru^2 dx \equiv 1; \quad (1)$$

where $a \neq b$, $u\left(\dfrac{du}{dx}\right) \equiv 0$ at a and b ; $p, q, r$ are functions of the real variable $x$.

The usual algorithm for succesively minimizing $\lambda$ generates increasingly higher degree polynomials. Letting $n$ denote the n[th] iteration, we obtain $\lambda_n < \lambda_{n-1}$ such that $\forall k \exists n [\lambda_{n-1} - \lambda_n < 1/k]$. We shall now call polynomials of sufficiently high degree (say, $k > 10^{50}$) an effective "eigenfunction". Every such effective eigenfunction, as it is a polynomial, is decomposable into "biunique eigenfunction pieces". We now show this theory is a foundation for physical fields governed by a nonlinear sigma model. Let us initially consider two effective eigenfunctions, $u_1(x_1)$ and $u_2(x_2)$; for each let $p \equiv 1$, $q \equiv 0$ and $r \equiv 1$ and we shall call $x_1$ "space" and $x_2$ "time". We note that in the theory T we can imply Hamilton's Principle for a one-dimensional string $u_1 u_2$ starting from an identity:

$$\int\left[\left(\frac{\partial u_1 u_2}{\partial x_1}\right)^2 - \left(\frac{\partial u_1 u_2}{\partial x_2}\right)^2\right] dx_1 dx_2 \equiv 0 \rightarrow \delta\int\left[\left(\frac{\partial u_1 u_2}{\partial x_1}\right)^2 - \left(\frac{\partial u_1 u_2}{\partial x_2}\right)^2\right] dx_1 dx_2 \equiv 0 \rightarrow \delta\int\left[\left(\frac{\partial u_1 u_2}{\partial x_1}\right)^2 - \left(\frac{\partial u_1 u_2}{\partial x_2}\right)^2\right] dx_1 dx_2 = 0$$

For each eigenstate *m*, we iterate the eigenfunctions $u_{1m}$ and $u_{2m}$ using the indicial expression $\lambda_{1m} \equiv \lambda_{2m}$. We immediately generalize to a Lagrange density for separable fields in finitely many space-like (i) and time-like (j) dimensions. As they are functions of real variables in T, the fields so obtained, or any finite sum of such fields, are locally homeomorphic, differentiable to all orders, of bounded variation and therefore without singularities.

Let $u_{\ell mi}(x_i)$ and $u_{\ell mj}(x_j)$ be effective eigenfunctions with non-negative eigenvalues $\lambda_{\ell mi}$ and $\lambda_{\ell mj}$ respectively. We assert a "field" is a sum of eigenstates $\underline{\Psi}_m = \sum_\ell \Psi_{\ell m} \underline{i}_\ell$, $\Psi_{\ell m} = \prod_i u_{\ell mi} \prod_j u_{\ell mj}$, subject to the postulate: <u>for every eigenstate *m* the integral of the Lagrange density over space-time is *identically* null.</u>

Let $ds d\tau = \prod_i r_i dx_i \prod_j r_j dx_j$,

$$\sum_\ell \int \left\{ \sum_i \frac{1}{r_i}\left[P_{\ell mi}\left(\frac{\partial \Psi_{\ell m}}{\partial x_i}\right)^2 - Q_{\ell mi} \Psi_{\ell m}^2\right] - \sum_j \frac{1}{r_j}\left[P_{\ell mj}\left(\frac{\partial \Psi_{\ell m}}{\partial x_j}\right)^2 - Q_{\ell mj} \Psi_{\ell m}^2\right]\right\} ds d\tau \equiv 0 \;\; \text{for all } m. \quad (2)$$

In this integral expression the *P* and *Q* can be real functions of any of the $x_i$ and $x_j$, thus of any $\Psi_{\ell m}$ as well. This is a *nonlinear sigma model*. The $\Psi_{\ell m}$ can be determined by coordinated iterations that are constrained by the indicial expression $\sum_{\ell i} \lambda_{\ell mi} - \sum_{\ell j} \lambda_{\ell mj} \equiv 0$ *for all m*. [7]

From expression (2) we can *derive quantization*. Since they are identical, we will represent both

$$\sum_{\ell mi} \int \left\{ \frac{1}{r_i}\left[P_{\ell mi}\left(\frac{\partial \Psi_{\ell m}}{\partial x_i}\right)^2 - Q_{\ell mi} \Psi_{\ell m}^2\right]\right\} ds d\tau \;\; \text{and} \;\; \sum_{\ell mj} \int \left\{ \frac{1}{r_j}\left[P_{\ell mj}\left(\frac{\partial \Psi_{\ell m}}{\partial x_j}\right)^2 - Q_{\ell mj} \Psi_{\ell m}^2\right]\right\} ds d\tau \;\; \text{by } \alpha\text{:}$$

I   $\alpha$ is positive and must be closed to addition and to the absolute value of subtraction;
    In the theory T, $\alpha$ must be an integer times a constant which is infinitesimal or finite.

II  If the field is not present, $\alpha \equiv 0$; otherwise, if the field is present, then in T $\alpha$ must be finite so that $\alpha \neq 0$; thus $\alpha = 0 \leftrightarrow \alpha \equiv 0$.

III  $\therefore \alpha \equiv n\iota$, where *n* is an integer and $\iota$ is a finite constant such that $\alpha = 0 \leftrightarrow n \equiv 0$.

If we attribute dimensions of action to $\iota$ then *the nonlinear sigma model expresses symmetry of action*. Expression (2) implies Hamilton's principle and, when there are finitely many space dimensions and one time-like dimension, we can obtain from it the Schrödinger equation: Let $\ell = 1, 2$ and suppress *m*; introduce $\Psi = \mathbf{A} \prod_i u_i(x_i)[u_1(\tau) + \mathbf{i} u_2(\tau)]$, where $\mathbf{i} = \sqrt{-1}$, and normalize $[u_1^2(\tau) + u_2^2(\tau)] \int \prod_i u_i^2(x_i)\, ds \equiv 1$.

We see that $\dfrac{du_1}{d\tau} = -u_2$ and $\dfrac{du_2}{d\tau} = u_1$ or $\dfrac{du_1}{d\tau} = u_2$ and $\dfrac{du_2}{d\tau} = -u_1$. For an *irreducible* time-eigenfunction the integral $\alpha$ for each and every biunique piece has its minimal non-zero value, i.e., the finite constant $\iota$. Thus $\mathbf{A}^2 \int \int_{n\pi/2}^{(n+1)\pi/2} \left[\left(\dfrac{du_1}{d\tau}\right)^2 + \left(\dfrac{du_2}{d\tau}\right)^2\right] \prod_i u_i^2(x_i)\, ds\, d\tau \equiv \mathbf{A}^2 \pi/2 \equiv \iota$. Letting $\iota$ be the Planck constant h/4 and letting $\tau$ be $\omega t$, the integrand becomes the usual time-part of the Lagrange density for the Schrödinger equation, $(h/4\pi\mathbf{i})\left[\Psi^*\left(\dfrac{\partial \Psi}{\partial t}\right) - \left(\dfrac{\partial \Psi^*}{\partial t}\right)\Psi\right]$. It immediately follows that the energy in the $m^{th}$ eigenstate, defined as this time-part of the Lagrange density, exists only in quanta of $h\omega_m/2\pi$. The sum of energies in all of the eigenstates $E_t$ is thus $\sum n_m h\omega_m/2\pi$ where $n_m$ is the number of quanta in the $m^{th}$ eigenstate.

We now offer a time-scale invariant argument for the definability in T of any field obtained from expression (2) in finitely many space-like dimensions and one time-like dimension in a finite region of space. Every ordered set $\{n_m\}$ corresponding to a distribution of a finite energy $E_t$ among eigenstates of this field maps to a unique finite integer and every finite integer maps to a unique ordered set $\{n_m\}$ by the fundamental theorem of arithmetic.

The set $\{n_m\} \Leftrightarrow$ the integer $\prod_m P_m^{n_m}$ where $P_m$ is the $m^{th}$ prime starting with 2. (3)

A set (in T) of these finite integers exists for all finite $E_t$. So quantization implies definability for a finite $E_t$. Moreover, in T we also can show that definability implies quantization. For finite $E_t$, if $\iota$ were infinitesimal, then $\sum n_m \omega_m$ would be infinite and the set of all distributions of energy among the eigenstates cannot be mirrored by a set (in T) of finite integers. Thus *definability in T is necessary and sufficient for quantization*.

In addition to providing a foundation for quantum mechanics, here are three examples of the applicability of theory T to physics. First, in T fields are constructed basically from smoothly connected biunique eigenfunction pieces. As we have seen, these pieces exist as a result of a homeomorphic mapping which is symmetric between range and domain. This construction necessitates not only that there are no discontinuities of the field but also that if the field vanishes the space must vanish, i.e., space-time is relational. Second, an important problem described by Dyson [8], that the perturbation series used in quantum electrodynamics are divergent, is absent in this theory, i.e., any induction in the theory must be finite, so perturbation series exist only to finite order and divergence cannot happen. Furthermore, all singularities that appear at the Fermi scale will be resolved at the Planck scale, since fields, as defined in T, can have no singularities. These examples suggest that theory T offers a possible foundation for quantum gravity. Third, the deep question raised by Wigner [9] concerning the unreasonable effectiveness of mathematics in physics is here answered directly.

Acknowledgement: The author's thanks to Vatche Sahakian, Maksim Perelstein and Harold Hodes of Cornell University and Rathin Sen of Ben-Gurion University for their advice and comments.

# Footnotes

1. Tarski, A., Mostowski, A., Robinson, R., *Undecidable Theories*. North Holland, Amsterdam, 1953.
2. Gödel, K., The consistency of the axiom of choice and of the generalized continuum hypothesis. *Annals of ath. Studies*, 1940.
3. Cohen, P. J., *Set Theory and the Continuum Hypothesis*, New York, 1966.
4. The axiom schema of subsets is $\exists u[[u = 0 \lor \exists x x \in u] \land \forall x x \in u \leftrightarrow x \in z \land X(x)]$, where $z$ is any set and $X(x)$ is any formula in which $x$ is free and $u$ is not free. The axiom enters ZF in AR but can also enter in the strong form of the axiom of regularity. (Note T has the weak form.) This axiom is essential to obtain the diagonal set for Cantor's proof, using $x \notin f(x)$ for $X(x)$, where $f(x)$ is an assumed one-to-one mapping between $\omega^*$ and the set of subsets of $\omega^*$. The argument leads to the contradiction $\exists c \in z X(c) \leftrightarrow \neg X(c)$, where $f(c)$ is the diagonal set. In ZF, this denies the mapping exists. In T, the same argument instead denies the existence of the diagonal set, which is not constructible and whose existence is an assumption while the mapping is asserted as an axiom. What if we tried the approach of using ABR to get a characteristic function? Let $\phi(x,y) \leftrightarrow [X(x) \leftrightarrow y = (x,1) \land \neg X(x) \leftrightarrow y = (x,0)]$ and $z = \omega^*$. If c were a member of $\omega^*$, $t = (c,1)$ and $t = (c,0)$ both lead to a contradiction. But, since the existence of the diagonal set $f(c)$ is denied and since a one-to-one mapping between the members of $\omega^*$ and the set of subsets of $\omega^*$ is an axiom, as $f(c)$ is not a subset of $\omega^*$, so c cannot be a member of $\omega^*$. In T the characteristic function exists but has no member corresponding to a diagonal set.
5. The axiom of constructibility generates sequentially all the subsets of $\omega^*$ in a set of ordered pairs. The left-hand member of each pair is a subset of $\omega^*$ and the right-hand member is an integer indicating the order in which it was generated. If we let the integers not present in each subset be a "1" in the corresponding binimal and the integers that are present be a "0", then the right-hand member is the magnitude of that binimal and serves as a distance measure on the line R*.
6. We have required the physical universe to be logical and this logic to be two-valued. Therefore physical behavior will be digital, that is, the universe becomes fundamentally a logic machine or "computer". Infinite series cannot be calculated *exactly* on a computer since we cannot have an infinite number of iterations.
7. The $u_{\ell m i}(x_i)$ and $u_{\ell m j}(x_j)$ are iterated using (1). The $p_{\ell m i}(x_i), q_{\ell m i}(x_i), p_{\ell m j}(x_j)$ and $q_{\ell m j}(x_j)$ will generally change at each iteration and are given by $p_{\ell m i} = \int \frac{P_{\ell m i} \Psi_{\ell m}^2 d\tau}{u_{\ell m i}^2 r_i dx_i} \bigg/ \int \frac{\Psi_{\ell m}^2 d\tau}{u_{\ell m i}^2 r_i dx_i}$ , etc.

Since the field is continuous, differentiable to all orders and of bounded variation, iterations for all $u_{\ell m i}(x_i)$ and $u_{\ell m j}(x_j)$, given the constraint $\sum_{\ell i} \lambda_{\ell m i} - \sum_{\ell j} \lambda_{\ell m j} \equiv 0$, will converge jointly within a finite region.

8. Dyson, F. J., Divergence of Perturbation Theory in Quantum Electrodynamics, Phys.Rev., 1952, 85.
9. Wigner, E. P., The Unreasonable Effectiveness of Mathematics in the Natural Sciences, Comm. Pure and      Appl. Math. 1960, 13.

# Appendix

## ZF – Subsets – Powers + Constructibility

Extensionality -   Two sets with just the same members are equal.

$$\forall x \forall y [\forall z [z \in x \leftrightarrow z \in y] \rightarrow x = y]$$

Pairs -   For every two sets, there is a set that contains just them.

$$\forall x \forall y \exists z [\forall w  w \in z \leftrightarrow w = x \vee w = y]$$

Union -   For every set of sets, there is a set with just all their members.

$$\forall x \exists y \forall z [z \in y \leftrightarrow \exists u [z \in u \wedge u \in x]]$$

Infinity -   There is at least one set with members determined in infinite succession

$$\exists \omega^* [0 \in \omega^* \wedge \forall x [x \in \omega^* \rightarrow x \cup \{x\} \in \omega^*]]$$

Regularity -   Every non-empty set has a minimal member (i.e. "weak" regularity).

$$\forall x [\exists y  y \in x \rightarrow \exists y [y \in x \wedge \forall z \neg [z \in x \wedge z \in y]]]$$

Replacement -   Replacing members of a set one-for-one creates a set (i.e., "bijective" replacement).

Let $\phi(x,y)$ a formula in which x and y are free,

$$\forall z \forall x \in z \exists y [\phi(x,y) \wedge \forall u \in z \forall v [\phi(u,v) \rightarrow u = x \leftrightarrow y = v]] \rightarrow \exists r \forall t [t \in r \leftrightarrow \exists s \in z \phi(s,t)]$$

Constructibility -   The subsets of ω* are countably constructible.

$$\forall \omega^* \exists S [(\omega^*, 0) \in S \wedge \forall y \forall z [\exists x  x \in y \wedge (y,z) \in S \leftrightarrow$$
$$\exists m_y [m_y \in y \wedge \forall v \neg [v \in y \wedge v \in m_y] \wedge \exists t_y [\forall u [u \in t_y \leftrightarrow u \in y \wedge u \neq m_y] \wedge (t_y \cup m_y, z \cup \{z\}) \in S]]]].$$